\documentclass[twocolumn,showpacs,preprintnumbers,prb,amsmath,amssymb]{revtex4}

\usepackage{graphicx}
\usepackage{dcolumn}
\usepackage{bm}
\usepackage{bbold}
\usepackage{booktabs}

\begin{document}

\title{High-precision finite-size scaling analysis of the quantum-critical \\ 
point of S\,=\,1/2 Heisenberg antiferromagnetic bilayers}

\author{Ling Wang}
\email{lingwang@buphy.bu.edu}

\author{K.\ S.\ D.\ Beach}
\email{ksdb@bu.edu}

\author{Anders W.\ Sandvik}%
\email{sandvik@bu.edu}

\affiliation{Department of Physics, Boston University, 
590 Commonwealth Avenue, Boston, Boston, Massachusetts 02215}

\date{\today}
             
\begin{abstract}
We use quantum Monte Carlo (stochastic series expansion) and finite-size scaling
to study the quantum critical points of two $S=1/2$ Heisenberg antiferromagnets 
in two dimensions: a bilayer and a Kondo-lattice-like system (incomplete bilayer),
each with intra- and inter-plane couplings $J$ and $J_\perp$. We discuss the 
ground-state finite-size scaling properties of three different quantities---the Binder moment 
ratio, the spin stiffness, and the long-wavelength magnetic susceptibility---which we 
use to extract the critical value of the coupling ratio $g=J_\perp/J$. The individual
estimates of $g_{\text{c}}$ are consistent provided that subleading finite-size corrections 
are properly taken into account. In the case of the complete bilayer, the Binder ratio leads 
to the most precise estimate of the critical coupling, although the subleading 
finite-size corrections to the stiffness are considerably smaller. For the incomplete 
bilayer, the subleading corrections to the stiffness are extremely small, and this 
quantity then gives the best estimate of the critical point. Contrary to predictions, 
we do not find a universal prefactor of the $\sim 1/L$ spin stiffness scaling at 
the critical point, whereas the Binder ratio is consistent with a universal value. 
Our results for the critical coupling ratios are $g_{\text{c}}=2.52181(3)$ (full bilayer) and 
$g_{\text{c}}=1.38882(2)$ (incomplete bilayer), which represent improvements of two 
orders of magnitude relative to the previous best estimates. For the correlation 
length exponent we obtain $\nu = 0.7106(9)$, consistent with the expected 3D Heisenberg 
universality class.
\end{abstract}


\maketitle

\section{\label{intro}Introduction}

The two-dimensional (2D) $S=1/2$ Heisenberg antiferromagnet has received considerable 
attention in the past two decades because of its close relation to the CuO$_2$ 
layers of the cuprate superconductors.\cite{ManousakisRMP1990} Other, even 
better realizations of this model system have been discovered as well.
\cite{RonnowPRL2001} The properties of the single-layer Heisenberg model have been 
thoroughly studied using both analytical and numerical methods, and there is 
very good agreement with experiments, \emph{e.g.}, for the temperature dependence 
of the spin correlation length \cite{ChakravartyPRL1988,KimPRL1998} 
in La$_2$CuO$_4$ (measured using neutron scattering) 
and NMR relaxation rates.\cite{SandvikPRB1995} 

Mapping the lattice Heisenberg model onto a continuum field theory yields the 
(2+1)-dimensional nonlinear sigma-model,\cite{HaldanePRL1983,ChakravartyPRL1988} 
the coupling constant $g$ of which controls the transition from N\'eel order to 
quantum disorder at temperature $T=0$ (quantum phase transition \cite{SachdevBOOK1999}).
This transition is in the universality class of the
finite-$T$ transition of the 3D classical Heisenberg model.
\cite{ChakravartyPRL1988,HaldanePRL1988} Having an ordered ground state, 
\cite{RegerPRB1988} the 2D square-lattice $S=1/2$ Heisenberg model corresponds to 
$g < g_{\text{c}}$. Even so, there is some influence from the critical point,
because a quantum phase transition is also associated with universal quantum 
critical scaling at finite temperature, in an extended ($g,T$) regime where 
temperature is the dominant energy scale.\cite{ChubukovPRB1994} The energy scales
characterizing the ordered and disordered phases---the spin stiffness and the 
singlet-triplet gap, respectively---vanish continuously as $g \to g_{\text{c}}$, and 
hence the quantum critical regime fans out from the point $(g=g_{\text{c}},T=0)$.

A quantum phase transition of the type described by the nonlinear sigma-model
can be realized in the Heisenberg antiferromagnet by introducing a pattern of 
two (or more) different coupling strengths in a way that favors
singlet formation on dimers (or larger units of an even number of spins).
\cite{ChakravartyPRL1988,ChubukovPRB1994} This leads to an order-disorder 
transition at some critical coupling ratio. Models of this kind, \emph{e.g.}, a bilayer 
where dimers form across the layers,\cite{HidaJPSJ1990,MillisPRL1993,SandvikPRL1994} 
single layers with various dimer patterns,\cite{SinghPRL1988} or a regularly depleted 
system where singlets form on rings of four or eight spins,\cite{TroyerPRL1996}  
have been extensively studied using quantum Monte Carlo simulations in order to 
confirm the expected universality class and to test very detailed predictions 
\cite{ChubukovPRB1994} of the finite-$T$ quantum critical behavior of various 
quantities. The predicted universal behavior was confirmed at low temperature.
\cite{SandvikPRL1994,ShevchenkoPRB2000,TroyerPRL1996,BrenigCM2005} The 
simulations also served to establish the onset of nonuniversal lattice 
effects at higher temperature and the nature of the cross-over 
\cite{ChubukovPRB1994} to the low-temperature renormalized classical or 
quantum-disordered regimes away from the critical point.

\begin{figure}
\includegraphics{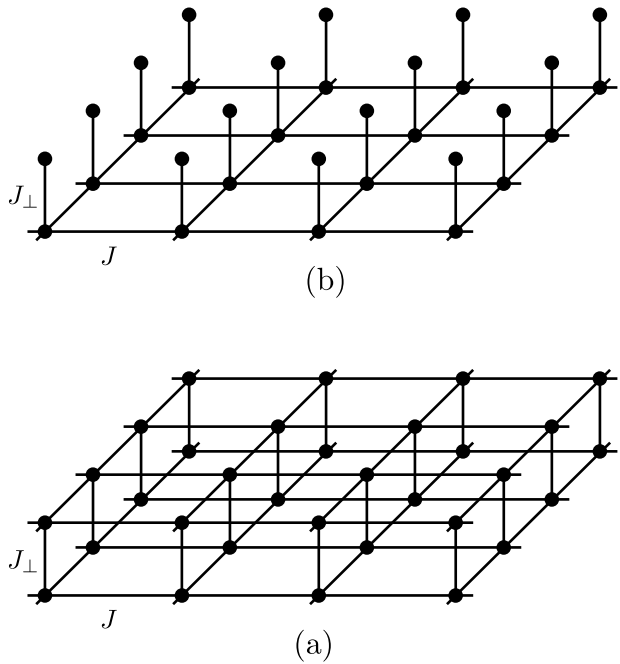}
\caption{The arrangement of spin interactions 
in the (a) full and (b) incomplete bilayers.
There are two different couplings: $J$ (intra-plane)
and $J_\perp$ (inter-plane), as indicated.}
\label{lattice}
\end{figure}

In this paper we study the critical points of two different $S=1/2$ Heisenberg
models: a symmetric bilayer and a Kondo-lattice-like system in which there are no 
intra-plane couplings in one of the layers. We will refer to these systems as the 
full and incomplete bilayers, respectively; see Fig.~\ref{lattice}. 
Their Hamiltonians ($H_1$ for the full
bilayer and $H_2$ for the incomplete bilayer) are given by
\begin{align}
\label{BlayerH}
H_1 & =  J\sum_{\langle i,j\rangle}\bigl({\bf S}_{1i}\cdot {\bf S}_{1j}+
{\bf S}_{2i}\cdot {\bf S}_{2j}\bigr)+J_{\perp}\sum_{i} {\bf S}_{1i}\cdot 
{\bf S}_{2i},\\
\label{DlayerH}
H_2 & =  J\sum_{\langle i,j\rangle} {\bf S}_{1i}\cdot {\bf S}_{1j}+
J_{\perp}\sum_{i} {\bf S}_{1i}\cdot {\bf S}_{2i}.
\end{align}
Here, ${\bf S}_{a,i}$ is a spin-1/2 operator at site $i$ of layer $a$ ($a=1,2$), 
and $\langle i,j\rangle$ denotes a pair of nearest-neighbor sites on the square
lattice of $L\times L$ sites with periodic boundary conditions. Both coupling constants 
are antiferromagnetic ($J,J_{\perp} > 0$). In order to avoid any frustration we
consider only even $L$. As the ratio $g=J_{\perp}/J$ is increased, there is a 
tendency to form inter-plane near-neighbor singlets, which at some $g=g_{\text{c}}$ 
leads to the opening of a spin gap and destruction of the long-range N\'eel order 
present for $g < g_{\text{c}}$ (in the limit $g \to \infty$ the ground state is a 
singlet product, which clearly cannot support long-range spin-spin correlations). 
This is the transition we investigate in detail in this paper.

Our purpose in studying these models is two-fold. First, we would like to determinate
the locations of the critical points to much higher accuracy than they are currently known.
The best estimates to date are $g_{\text{c}} = 2.525(2)$ (full bilayer)
\cite{ShevchenkoPRB2000} and $g_{\text{c}} = 1.393(8)$ (incomplete bilayer)
\cite{KotovPRL98} (see also Ref.~\onlinecite{BrenigCM2005}). The statistical 
accuracies here are quite modest compared to results for the standard classical 
critical points (\emph{e.g.}, the 3D Heisenberg model \cite{HolmPRB1993,CompostriniPRB2002}). 
It would be useful to increase the precision so that studies of various aspects of 
finite-$T$ quantum criticality (\emph{e.g.}, interesting properties of isolated  impurities 
in a critical host system 
\cite{SachdevSCI1999,TroyerPTP2002,SachdevPRB2003,SushkovPRB2003,HoglundUNP}) 
could be studied numerically at low $T$ very close to the critical point (minimizing the
effects of the eventual cross-over to the renormalized-classical or 
quantum-diosrdered regime). Second, we wish to compare several different ways 
of extracting the critical coupling, in order to gain additional confidence in the 
results and to provide guidance for studies of other quantum critical points. The 
reason for choosing the particular bilayer lattices of Fig.~\ref{lattice}
over other 2D Heisenberg systems undergoing the same type of transition is that 
they do not break any in-plane symmetries of the square lattice.

We have carried out finite-size scaling of low-temperature ($T\to 0$ converged)
QMC results for three different quantities: the spin stiffness, the Binder
cumulant ratio, and the long-wavelength (uniform) magnetic susceptibility. 
We use our recently proposed method for including subleading finite-size
corrections.\cite{KevinPRE05} Although this necessitates nonlinear fits with 
a relatively large number of independent parameters, we believe that this is 
necessary in order to obtain completely unbiased results. Our final results for 
the critical couplings are $g_{\text{c}} = 2.52181(3)$ for the full bilayer and 
$g_{\text{c}} = 1.38882(2)$ for the incomplete bilayer, \emph{i.e.}, the statistical precision 
is improved by approximately two orders of magnitude relative to the previous 
estimates. Our fitting procedure also gives the correlation-length exponent 
$\nu$, but because of the multi-parameter fits and the relatively modest 
lattices sizes ($L$ up to $42$), its statistical precision is not quite as high 
(the error bars are roughly twice as large) as that of recent classical Monte Carlo 
simulations of the 3D Heisenberg model.\cite{CompostriniPRB2002} 
Nevertheless, our result, $\nu = 0.7106(9)$, is fully consistent with 
the presently most accurate value of this exponent, $\nu=0.7112(5)$.
\cite{CompostriniPRB2002}

We also discuss the universality of the Binder moment ratio and the spin stiffness 
at the critical point. In the former case, we point out a difference relative to 
classical systems in how the order-parameter moments are defined and calculated, 
and in the latter case we do not find the universality that has been suggested for 
the prefactor of the $\sim 1/L$ scaling\cite{WallinPRB1994} (\emph{i.e.}, we obtain 
different prefactors for the two different bilayer models).

The rest of the paper is organized as follows. In Sec.~\ref{scaling} we
discuss the quantities that we have calculated and their QMC (Stochastic
Series Expansion, SSE) estimators, as well as their expected finite-size 
scaling forms. In Sec.~\ref{data} we first briefly review our approach to 
deal with  subleading finite-size corrections and then present the results 
of the analysis. We give a brief summary and conclusions in Sec.~\ref{summary}.

\section{\label{scaling}Calculated observables and their
critical scaling properties}

\begin{figure}
\includegraphics{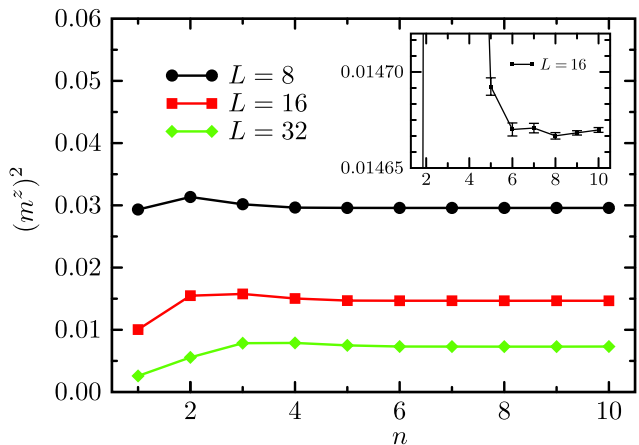}
\caption{(Color online) Convergence as a function of inverse temperature 
$\beta=2^n$ of the squared sublattice magnetization to its ground state value 
for three different lattice sizes.}
\label{betaconv}
\end{figure}

We have used the SSE QMC method with operator-loop updates.\cite{SandvikPRB1999} 
This approach is based on sampling diagonal matrix elements of the power series 
expansion of $\exp(-\beta H)$, where $\beta$ is the inverse temperature. We use
$L\times L\times 2$ lattices with periodic boundary conditions in
the $x$ and $y$ directions, with even $L$ up to $42$. In order to ensure 
convergence of all calculated quantities to their ground-state values, we carried 
out simulations at inverse temperatures $\beta_n =2^n$ with integer $n$ taken
large enough so that results for $\beta_{n}$ and $\beta_{n-1}$ agree within 
statistical errors. Examples of the convergence are shown in Fig.~\ref{betaconv}.

\subsection{\label{q2}Binder Moment Ratios $Q_k$}

The magnetic moment ratios $Q_k$ introduced by Binder~\cite{BinderPRL1981}
have the very useful property of being universal at the critical point, because 
all nonuniversal scale factors cancel out along with the length dependence.
This follows from the finite-size scaling hypothesis for the ordered
moment (here the staggered magnetization). The $k^{\text{th}}$ power of the
staggered magnetization scales as
\begin{equation}
\label{spinmoments}
{\langle |m|^k\rangle}_L=L^{-k\beta/\nu}M_k(tL^{1/\nu}),
\end{equation}
where $L$ is the linear system length, $\beta$ and $\nu$ are critical
indices in their standard notation, and $t$ is the reduced coupling constant, 
which we define here in terms of the coupling ratio $g$ as $t=(g-g_{\text{c}})/g_{\text{c}}$. 
$M_k(x)$ are the scaling functions. Consequently, the moment ratios 
\begin{equation}
\label{scaleq2}
Q_k(t,L)=\frac{{\langle m^{2k}\rangle}_L}{{\langle m^2\rangle}^k_L}
\end{equation}
are dimensionless scaling functions. At the critical point,
$Q_k(0,\infty)$ are universal constants.

We have computed the first two Binder ratios, which we
define as
\begin{eqnarray}
Q_1 & = & \frac{\langle m^2\rangle}{{\langle |m|\rangle}^2} 
=\frac{3\langle(m^z)^2\rangle}{2{\langle |m^z|\rangle}^2}, 
\label{measq1} \\
Q_2 & = &\frac{\langle m^4\rangle}{{\langle m^2\rangle}^2}
=\frac{5\langle (m^z)^4\rangle}{3{\langle(m^z)^2\rangle}^2},
\label{measq2}
\end{eqnarray}
where $m^z$ is the $z$-component of the staggered magnetization 
operator,
\begin{equation}
m^z = \frac{1}{N}\sum_{i=1}^{N}S_i^z(-1)^{x_i+y_i} = m \cos(\Theta).
\label{mz}
\end{equation}
Here,  $N=2L^2$ is the number of lattice sites. Since the $O(3)$ spin-rotational
symmetry is not broken in the simulations (\emph{i.e.}, an average over all angles
$\Theta$ is obtained) we have included the appropriate factors to compensate for 
the rotational averaging of $m^z$ in Eqs.~(\ref{measq1}) and (\ref{measq2}).  

\begin{figure}
\includegraphics{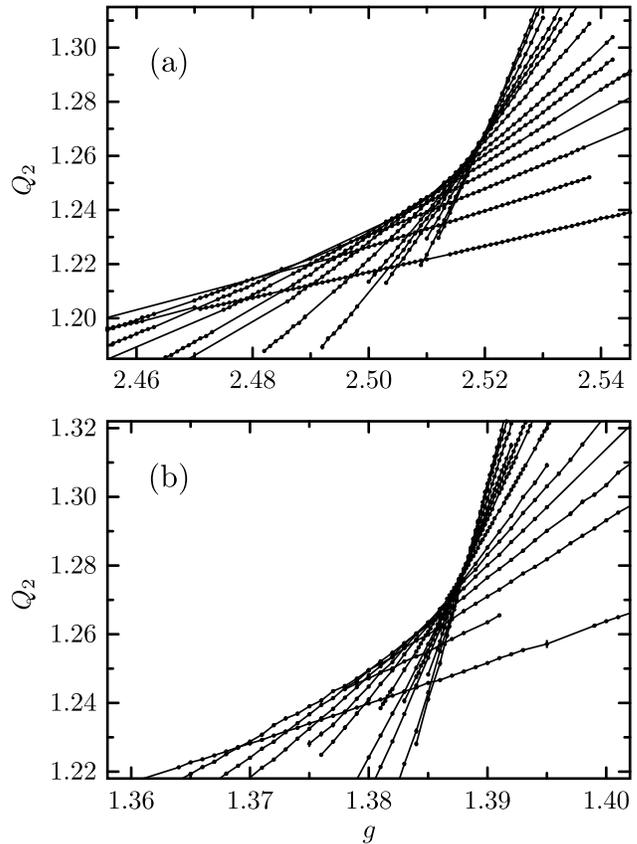}
\caption{\label{fig:q2-crss}
Binder ratio $Q_2$ for different system sizes vs the coupling ratio for 
the full (a) and incomplete (b) bilayers. Results for even $L$ from $8$ to $42$ 
are shown (all except for $L=22,26,34,38$). The slope of the curves
increases with $L$.}
\end{figure}

In Fig.~\ref{fig:q2-crss} we show the ratio $Q_2$ for both bilayer
systems as a function of $g$ for lattices of different linear length $L$. 
One can clearly see the curve crossings, indicating a quantum critical 
point, but it is apparent that there are sizable corrections to their location.
We will analyze these crossing-point shifts in Sec.~III.

Note that $Q_2$ at the crossing point is approximately the same for both 
models, in accord with an expected universal value as $L\to \infty$. Simulations 
done on 3D classical Heisenberg models \cite{HolmPRB1993,CompostriniPRB2002,CaparicaPRB2000} 
gave a universal value in the range 1.35--1.40, \emph{i.e.}, substantially larger than 
what we see in Fig.~\ref{fig:q2-crss} (clearly these values have not yet completely
converged to their infinite-size $Q_2$, and the trend is for the crossing value 
to increase with $L$, but we will show that they converge to $Q_2\approx 1.29$). 
This disagreement with the classical value is easily accounted for by considering
the way the sublattice magnetization is defined and computed in a quantum system: 
Although the 2D system formally is mapped onto a 3D classical model, 
$\langle |m^z|^k\rangle$ is an equal-time expectation value, which in the
simulations is averaged over the third (imaginary time) direction. This corresponds 
directly to taking expectation values over individual layers in a 3D classical 
model. We are not aware of any such calculation and hence cannot compare directly 
with the corresponding classical universal value. Nevertheless, as we will show
in greater detail in Sec.~III, the crossing $Q_2$ values for both our systems 
are fully consistent with each other and hence support universality.

\subsection{\label{rho}Spin Stiffness $\rho_{\text{s}}$}

In continuum field theory language, a stiffness $\rho$ is defined in 
terms of the increase in free-energy density $f$ as a boundary-condition
twist $\Phi$ is imposed on the order-parameter field $\theta$:
\begin{equation}
\label{deltaf}
\delta f_s(t,L)=\frac{1}{2}\rho(\nabla\theta)^2
=\frac{1}{2}\rho(\Phi/L)^2.
\end{equation}
The prefactor is the stiffness constant,
\begin{equation}
\rho=L^2\frac{\partial^2f_s}{\partial\Phi^2}.
\end{equation}
At a quantum critical point, it should scale as \cite{FisherPRB1989}
\begin{equation}
\rho \sim L^{2-d-z},
\label{rhoscale}
\end{equation}
where $d$ is the dimensionality and $z$ is the dynamic critical exponent. 
Wallin  {\it et al.} have argued that $\rho L^{2-d-z}$ is a universal
number in two dimensions.\cite{WallinPRB1994}

\begin{figure}
\includegraphics{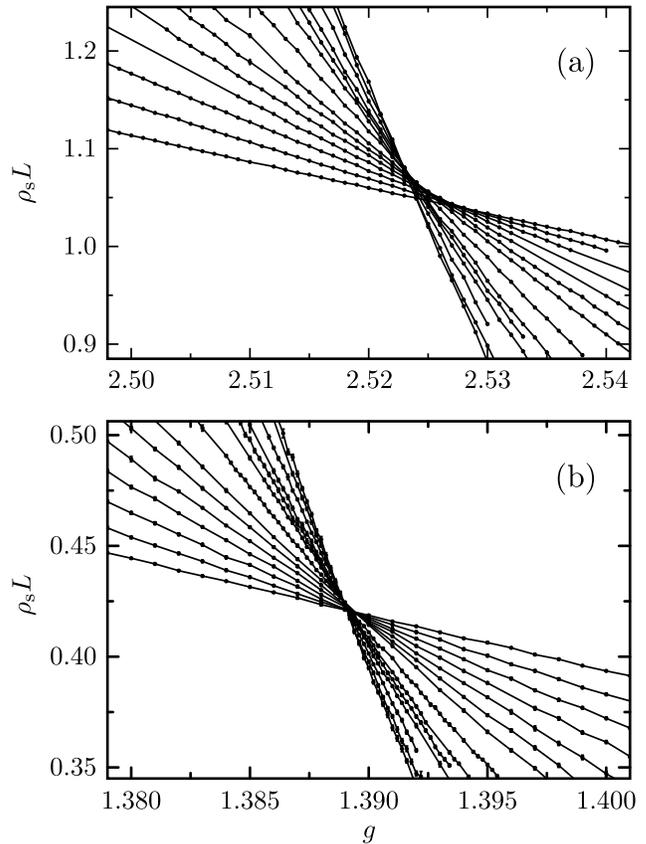}
\caption{\label{fig:rho-crss}
The spin stiffness multiplied by the system length $L$ at the critical 
point. The system sizes are the same as in Fig.~\ref{fig:q2-crss}. The 
slope of the curves increases with $L$.}
\end{figure}

For the Heisenberg model, the spin stiffness $\rho_{\text{s}}$ is determined by imposing 
a twist directly in the Hamiltonian, modifying the spin-spin interactions in one 
of the lattice directions: ${\bf S}_i \cdot {\bf S}_j$ $\to $ ${\bf S}_i \cdot 
{\bf R}(\Phi/L){\bf S}_j$, where ${\bf R}$ rotates the spin operator about an 
appropriately chosen axis.\cite{SandvikPRB1997} In SSE simulations, like in path 
integrals,\cite{PollockPRB1987} the stiffness is directly obtained without
explicitly imposing a twist, as the second derivative of the energy with respect to 
the twist at $\Phi=0$. This leads to an estimator in terms of  winding number 
fluctuations,\cite{SandvikPRB1997}
\begin{equation}
\rho_{\text{s}}=\frac{3}{4}\langle w_x^2+w_y^2\rangle /\beta,
  \label{rhomeas}
\end{equation}
where the winding numbers are
\begin{equation}
w_{\alpha}=(N_{\alpha}^+-N_{\alpha}^-)/L, \qquad (\alpha=x,y).
\end{equation}
Here $N_{\alpha}^+$ ($N_{\alpha}^-$) is the number of operators
$S_i^+S_j^-$ ($S_i^-S_j^+$) in the sampled terms of the power series 
expansion, with $i,j$ two nearest-neighbor sites oriented along the 
lattice $\alpha$ ($x$ or $y$) axis. The definition (\ref{rhomeas}) corresponds 
to the stiffness per unit cell of the bilayer models.

In the case of the bilayer models we have $d=2$ and expect $z=1$, and hence the 
scaling (\ref{rhoscale}) becomes $\rho_{\text{s}} \sim L^{-1}$. The quantity $\rho_{\text{s}}L$ should 
thus be size-independent at the critical point, and also in this case one can 
expect curves for different $L$ to cross each other. Such crossings have 
previously been used to approximately locate the critical point of the full 
bilayer.\cite{SandvikPRB1997} 

Fig.~\ref{fig:rho-crss} shows our SSE results for $\rho_{\text{s}}L$ versus $g$ for 
different lattice sizes. Again, one can see that the crossing points move as 
$L$ is increased, but, interestingly,  much less so for the incomplete than 
the complete bilayer. These results do not immediately support a universal 
$L\rho_{\text{s}}$ at $g_{\text{c}}$; a careful finite-size scaling analysis does not
change this conclusion, as we will see in Sec.~III.

\subsection{\label{xq0}Uniform Susceptibility $\chi(q\to 0)$}

The temperature dependence of the uniform magnetic susceptibility is an 
often-used indicator of quantum criticality. Exactly at $g=g_{\text{c}}$, the general
asymptotic scaling  form is
\cite{ChubukovPRB1994} 
\begin{equation}
\chi_u (T) \sim T^{d/z-1}.
\label{xuTscaling}
\end{equation}
This has been numerically confirmed at low $T$ in previous QMC simulations 
\cite{SandvikPRL1994,ShevchenkoPRB2000} and series expansions
\cite{ElstnerPRB1998} of the bilayer and other critical Heisenberg systems.
\cite{TroyerPRB1997} Here we will consider the corresponding finite-size scaling 
behavior, which we obtain by substituting the finite-$T$ quantum critical
correlation length, $\xi\sim T^{-1/z}$,\cite{ChakravartyPRL1988,ChubukovPRB1994} 
in the temperature dependence (\ref{xuTscaling}) and then substitute $L$ for $\xi$, 
giving $\chi_u \sim L^{z-d}$. However, we apparently have a problem here since the
uniform susceptibility $\chi_u=\beta\langle (M^z)^2\rangle$ vanishes as $T \to 0$, 
due to the conserved magnetization $M^z$ and the singlet ground state. In order to 
carry out finite-size scaling, we therefore consider the long-wavelength limit of 
the wave-vector dependent susceptibility $\chi({\bf q})$, which we obtain in practice
by taking $q=2\pi/L$. Thus we will test the finite-size scaling form 
\begin{equation}
\chi (q\to 0) = \chi(2\pi/L) \sim L^{z-d}.
\end{equation}

The static spin-spin susceptibility in real space is given by the
Kubo integral
\begin{equation}
\chi(k,l)=\int_0^{\beta}\!d\tau\,\bigl\langle S_k^z(\tau)S_l^z(0)\bigr\rangle ,
\end{equation}
which in SSE simulations is obtained in terms of spins in the states 
propagated by the sampled operator sequences:\cite{SandvikPRB1997}
\begin{multline}
\chi(k,l) = \Biggl\langle\frac{\beta}{n(n+1)}\left(\sum_{p=0}^{n-1}S_k^z[p]\right)
\left(\sum_{p=0}^{n-1}S_l^z[p]\right)\\
+\frac{\beta}{(n+1)^2}\left(\sum_{p=0}^{n}S_k^z[p]S_l^z[p]\right)
\Biggr\rangle.
\end{multline}
$n$ is the number of hamiltonian operators in the sampled sequences
and the index $p$ refers to the state obtained after $p$ operators have
acted. The Fourier transform that we are interested in is
\begin{equation}
\chi({\bf q})=\frac{1}{N}\sum_{k,l}e^{i{\bf q}\cdot({\bf r}_k-{\bf r}_l)}\chi(k,l),
\label{fourierchi}
\end{equation}
and since our models are symmetric with respect to a $90^\circ$ rotation,
we take the long-wavelength susceptibility as
\begin{equation}
\chi_u = \chi(q\to 0)=\frac{1}{2}\Bigl[\chi(\tfrac{2\pi}{L},0,0)+
\chi(0,\tfrac{2\pi}{L},0)\Bigr].
\end{equation}

\begin{figure}
\includegraphics{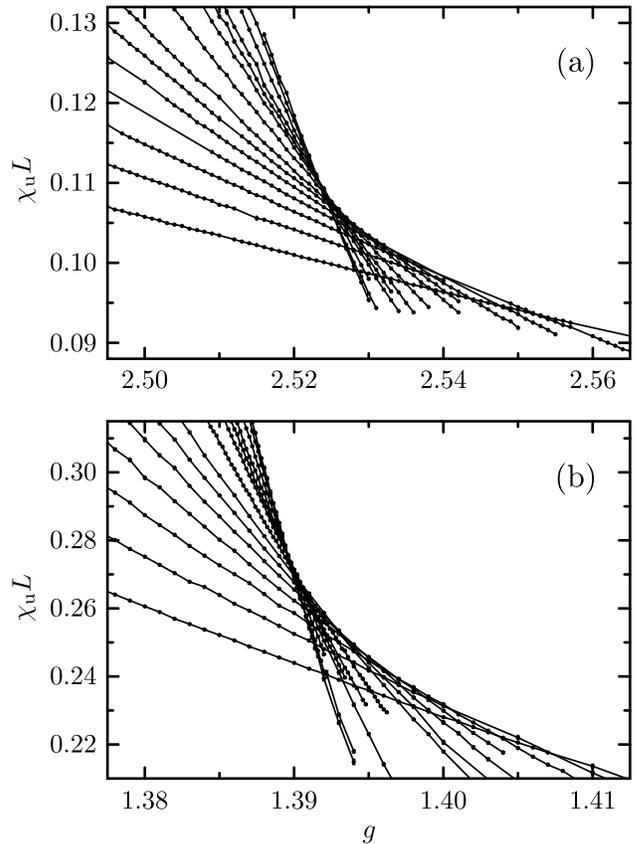}
\caption{\label{fig:xq0-crss}
The long-wavelength susceptibility versus the coupling ratio
for (a) the full bilayer and (b) the incomplete bilayer. The
system sizes are those listed in Fig.~\ref{fig:q2-crss}. The slope
of the curves again increases with $L$.}
\end{figure}

We again consider the form leading to curve crossings at the critical point, 
\emph{i.e.}, we plot $\chi_u L$, which should be size-independent at $g_{\text{c}}$. 
Figure~\ref{fig:xq0-crss} shows the data that we will anayze more
carefully in the next section. Again we observe crossing points, which
shift significantly with $L$.

\section{\label{data}Data Analysis}

We first discuss here a rough determination of the critical coupling ratios of 
the two models, studying the asymptotic behavior of the crossing points
discussed above. This will also serve as a first confirmation of mutual 
consistency of the leading scaling forms for the three different quantities
under consideration. We then analyze the results in greater detail using a 
finite-size scaling hypothesis including subleading corrections.

\subsection{Critical Coupling from Crossing Points}

\begin{figure}
\includegraphics{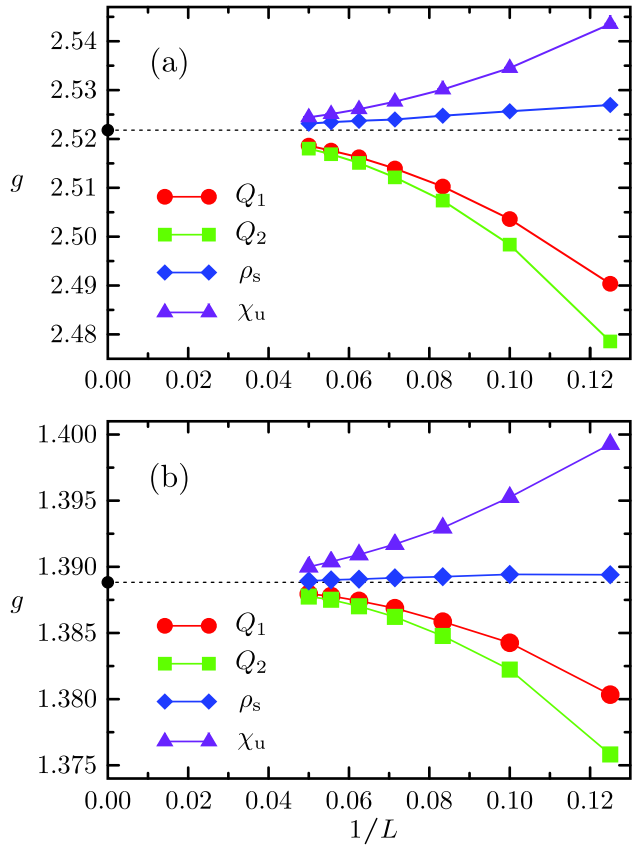}
\caption{\label{fig:cnvrg}(Color online) 
Convergence vs the inverse lattice size of intersection points of curves for 
$L$ and $2L$, for the full (a) and incomplete (b) bilayers. The error bars are
are much smaller than the symbols. The circles at $1/L=0$ indicate the critical 
couplings from the careful finite-size scaling analysis carried out in Sec.~III B.}
\end{figure}

We use the data presented in the previous 
section to extract the intersection points of fixed-$L$ curves for system sizes 
$L$ and $2L$ (other size ratios give similar results). Our 
simulations have been performed on a rather dense grid of $g$-points, and we 
can therefore reliably obtain the intersection points using fits of straight
lines or second-order polynomials to interpolate between the data points. 
In Fig.~\ref{fig:cnvrg} we plot the results versus the inverse system size. 
For both models, the crossing points drift toward a common critical coupling in the $L\to\infty$ limit,
thus confirming the scaling laws discussed in the prevcious section.

For both models, especially the incomplete bilayer, the spin stiffness 
crossing point exhibits the most rapid convergence (\emph{i.e.}, the weakest
subleading corrections). It and the susceptibility 
converge from above, while the Binder ratios converge from below. We can hence 
bracket $g_{\text{c}}$ using these results. However, a much more precise bracketing can be 
obtained from the spin stiffness curves alone, noting that they  become very flat 
as $L$ grows. With the curvature decreasing with increasing $L$, a straight-line 
extrapolation using a few large-$L$ points (we use four) should give a lower 
bound for $g_{\text{c}}$, while the crossing point for the largest $L$ should be an upper 
bound. The critical couplings extracted this way are $g_{\text{c}} \in(2.5205,2.5232)$ 
and $g_{\text{c}}\in(1.38870,1.38895)$ for the full and incomplete bilayers, respectively. 
These values are fully consistent with the best previous estimates, discussed 
in Sec.~I, but the precision is significantly higher. The more rigorous 
data analysis discussed below will further improve on the accuracy.

Naively, one might expect that the asymptotic approach of the crossing points 
to the critical coupling should be given by the correlation-length exponent $\nu$,
as $g_{\rm cross} = g_{\text{c}} +aL^{-1/\nu}$, 
as is the case for fixed-size estimates of the critical
coupling (or the critical temperature), such as the location of the maximum
of the order-parameter susceptibility or the specific heat (in the case
of finite-$T$ transitions). However, a crossing point cannot be regarded
as a conventional fixed-$L$ definition of $g_{\text{c}}$, since two system sizes are
involved and there can be cancelations of a leading behavior defined in terms
of the individual lattices. Thus, we would in general expect the crossing points to 
converge faster than $L^{-1/\nu}$. Binder  has discussed the corrections to the 
cumulant crossing-points,\cite{BinderPRL1981}
and in a recent article \cite{KevinPRE05} we have presented a different way 
of analyzing crossing points in general (\emph{i.e.}, not only for Binder ratios but 
also for other quantities that are size-independent at $g_{\text{c}}$) which takes
subleading finite-size corrections into account explicitly. There we also showed
some results for the spin stiffness crossings of the full bilayer model. Here 
we will not analyze the crossing points in any greater detail, but instead consider 
the scaling of the full fixed-$L$ curves shown in the figures of Sec.~II. 
Such ``data collapse'' makes better use of the full range of simulation results and 
can also be carried out with subleading corrections taken into account.\cite{KevinPRE05}

\subsection{Finite-Size Scaling with Subleading Corrections}

The scaling ansatz typically used to analyze finite-size data
$A(t,L)$ for a quantity $A$ at reduced coupling $t=(g-g_{\text{c}})/g_{\text{c}}$ on 
a lattice of length $L$ is
\begin{equation}
\label{collapse1}
A(t,L)=L^{\kappa/\nu}f_A(tL^{1/\nu}),
\end{equation}
where $\kappa$ is a critical exponent which depends on the quantity $A$,
\emph{i.e.}, $A(t,L=\infty) \sim  t^{-\kappa}$. This form can be used to collapse 
data in a neighborhood of $t=0$, by graphing $A(t,L)L^{\kappa/\nu}$ versus 
$tL^{-1/\nu}$, adjusting $g_\text{c}$, $\kappa$, and $\nu$ to obtain the tightest
collapse of the data onto a single curve. 

\begin{figure}[!t]
\includegraphics{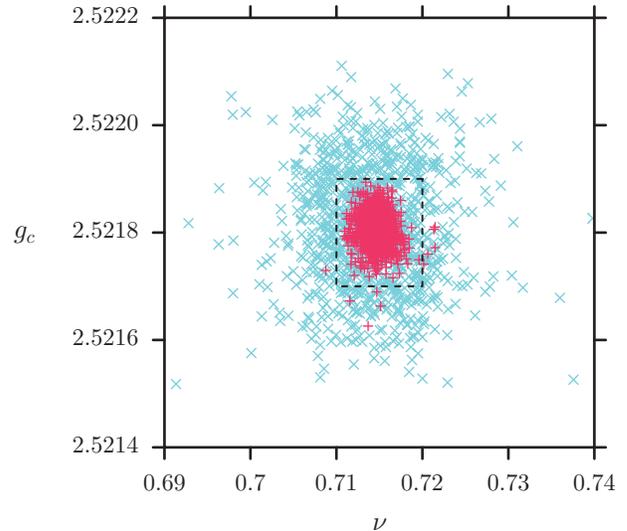}
\caption{ \label{FIG:Q2complete} (Color online) The blue crosses $(\times)$ show the
  input points of the optimization procedure. The red plusses $(+)$
  show the output.  The dashed black line indicates the magnified
  region correspondig to the bottom-right plot in
  Fig.~\ref{FIG:densityQ2rho}. }
\end{figure}

\begin{figure*}[!t]
\includegraphics{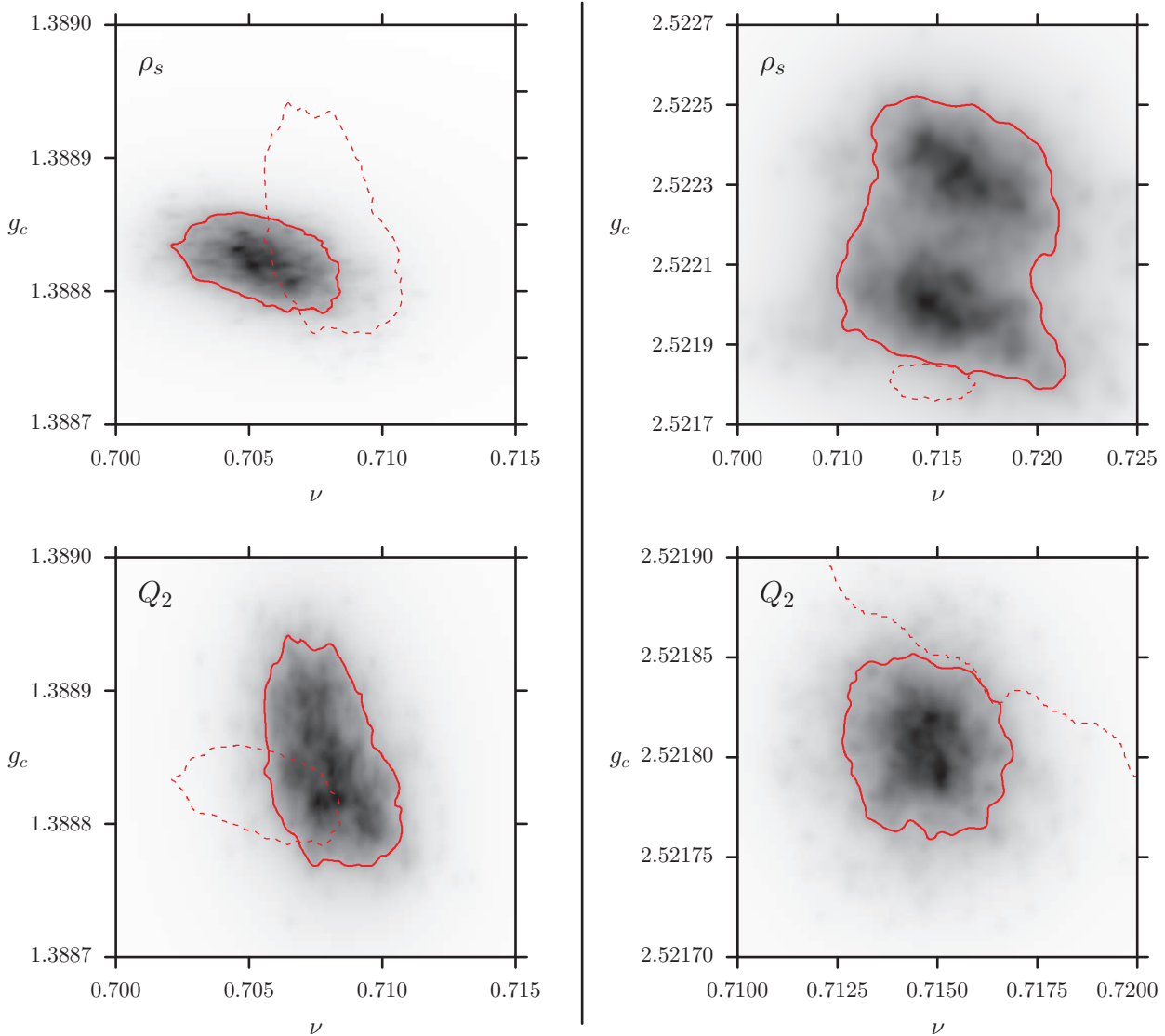}
\caption{ \label{FIG:densityQ2rho} (Color online)
  The number density of bootstrapped
  $(\nu,g_{\text{c}})$ solutions is plotted for $Q_2$ and $\rho_{\text{s}}$ in greyscale. 
  The left panel shows the incomplete bilayer and the right
  panel the complete bilayer.  The solid red lines show the contour at
  1/3 the maximum density (this would correspond to 2/3 of the weight
  under a gaussian distribution). In the top right panel, note a two-peak 
  structure which sometimes appears in the nonlinear fits. The dotted lines 
  are guides to the eye.  They show the $Q_2$ contour on the $\rho_{\text{s}}$ 
  plot and vice versa.  }
\end{figure*}

In Ref.~\onlinecite{KevinPRE05} 
we started from renormalization group theory and 
derived an extension to (\ref{collapse1})  that includes
both ``shift'' and ``renormalization'' corrections:
\begin{equation}
\label{collapse}
A(t,L)=L^{\kappa/\nu}(1+cL^{-\omega})g_A(tL^{1/\nu}+dL^{-\phi/\nu}).
\end{equation}
Here, $\phi$ is  the subleading irrelevant RG eigenvalue, which causes
a shift in the critical coupling. $\omega$ is an effective exponent that
accounts for corrections due to the inhomogeneous 
part of the free energy and nonlinearity of the scaling fields.
The constants $c$ and $d$ are nonuniversal and should
be regarded as fitting parameters along with the leading and subleading exponents. 
From Eq.~(\ref{collapse}), we see that we can now achieve data collapse by 
plotting $A(t,L)L^{-\kappa/\nu}/(1+cL^{-\omega})$ versus $x=tL^{1/\nu}+dL^{-\phi/\nu}$ 
for different sizes $L$. 

To carry out this type of analysis in practice, 
we note that the scaling function $g_A(x)$ is well-behaved and can be Taylor 
expanded close to the critical point:
\begin{eqnarray}
g_A(x)&=&A(t,L)L^{-\kappa/\nu}/(1+cL^{-\omega}) \label{taylor} \\
\label{gafunction}
\nonumber
&=&q_0+q_1x+q_2x^2+q_3x^3+q_4x^4+\cdots
\end{eqnarray}
For the Heisenberg bilayers, the  critical indices $\kappa$ and $\nu$ are 
expected to be those of the 3D classical Heisenberg universality class. In the
case of the ratios we are considering, $\kappa/\nu=$ are known integers which
we hence do not have to adjust. The current best estimate for the correlation 
length exponent is $\nu=0.7112(5)$,\cite{CompostriniPRB2002} 
but in our analysis we keep it as a free parameter, 
along with $g_{\text{c}}$, the subleading exponents $\omega$
and $\phi$, the constants $c$ and $d$, and the parameters of the
polynomial in (\ref{taylor}). This amounts to a large number of fitting 
parameters, but it should be noted that the polynomial expansion of the scaling
function is essentially just an interpolation of the data and hence
is highly constrained; the freedom in the coefficients $q_i$ do not
add significant freedom to the other paramers of the fit (we use a quartic
polynomial). Moreover, the number of fitting parameters is dwarfed
by the number of data points to be fit (hundreds or thousands). 

Nonlinear curve fitting has well-known problems associated
with the convergence of the parameters to the globally optimal fit. 
In our work we already 
know rather accurate estimates for $\nu$ and $g_{\text{c}}$, and at the first stage
of the fits we used those values as initial conditions. Once we obtained 
rough estimates for $c,\omega,d,\phi$, we used the following procedure: 
Performing bootstrap sampling of the raw data, we carried out a large number 
(typically around 1000) of fits with initial conditions for all the parameters 
taken at random from inside a ``box'' in parameter space. This box is determined 
such that the fits converge well, but that the variation in starting points 
is significantly larger than the final spread of the resulting parameters. We 
then use the spread among the bootstrap samples to calculate statistical errors. 
This procedure is illustrated in  Figs.~\ref{FIG:Q2complete} and
\ref{FIG:densityQ2rho}.

The scaling formula, Eq.~(\ref{collapse}), is strictly valid only for a 
small range of couplings in the vicinity of $g_{\text{c}}$ [although the range of validity 
should be larger than with the leading-order form (\ref{collapse1})]. 
The parameters obtained show some dependence on the range of data included. 
In order to eliminate as much as possible potential remaining effects of further 
subleading corrections that are not captured by our approach, we chose to use a 
rather narrow window in the scaled coupling $x=(g-g_{\text{c}})L^{1/\nu}/g_{\text{c}}+dL^{-\phi/\nu}$, 
so that there is no longer any statistically detectable dependence on the size 
of the window. Our final results are based on $x\in[-0.25,0.25]$. There are
also other subtle issues in the fitting procedure, \emph{e.g.}, for a given range 
of $x$, different number of data points are available for the different 
lattice sizes, typically leading to relatively smaller statistical weight
for the larger sizes than the smaller sizes. We therefore made sure to
include only system sizes sufficiently large for the extracted parameters not 
to change appreciably when systematically excluding more of the smaller lattices. 
We kept only $L \ge 8$ for the results we report here.

\begin{figure}
\includegraphics{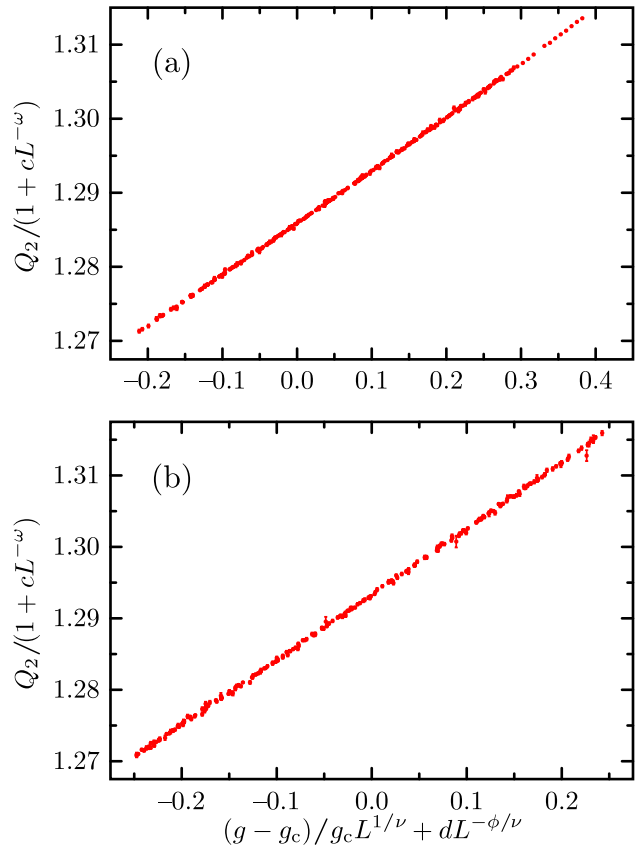}
\caption{\label{fig:q2-cllps}(Color online)
Data collapse of the Binder ratio for the full (a) and incomplete (b) bilayer.
The values of $g_{\text{c}}$ and $\nu$ obtained are listed in Table~\ref{tab:result}.}
\end{figure}

\begin{figure}
\includegraphics{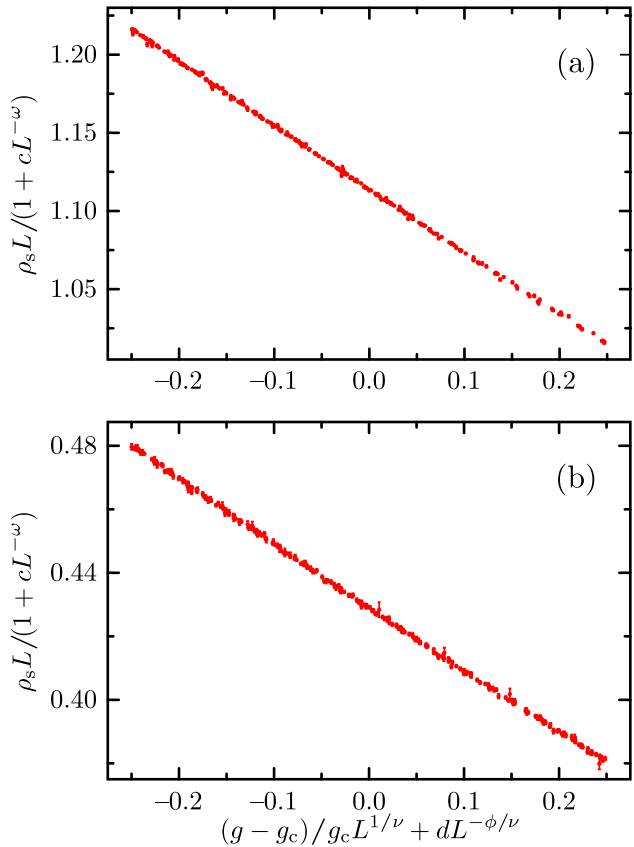}
\caption{\label{fig:rho-cllps}(Color online)
Data collapse of the spin stiffness ratio for the full (a) and incomplete (b) 
bilayer. The values of $g_{\text{c}}$ and $\nu$ obtained are listed in Table~\ref{tab:result}.}
\end{figure}

We found both the Binder ratio and the spin stiffness to be well behaved
in the fitting procedures. The resulting collapsed data for these quantities, 
\emph{i.e.}, their scaling functions $g_A(x)$ in Eq.~(\ref{gafunction}), are shown in 
Figs.~\ref{fig:q2-cllps} and \ref{fig:rho-cllps}. We do not discuss 
results for $Q_1$ here as it behaves similarly to $Q_2$ and is 
statistically strongly correlated to it. The long-wavelength susceptibility 
also exhibits data collapse, however with substantially larger fluctuations 
than $\rho_{\text{s}}$ and $Q_2$ (the prefactor $d$ appears to be rather
large and difficult to determine accurately, and thus it is difficult
to fix the data window $x \in [x_{\rm min},x_{\rm max}]$ in a meaningful way). 
We have therefore focused on $\rho_{\text{s}}$ and $Q_2$ for the final high-precision 
statistical analysis.

The final parameters and their statistical errors were determined
from the bootstrap samples. The distributions are illustrated 
by density plots in Fig.~\ref{FIG:densityQ2rho}. We list the
values for $g_{\text{c}}$, $\nu$, and the value of the respective quantities at
the critical point, $q_c=Q_2(g_{\text{c}}),\rho_{\text{s}}(g_{\text{c}})L$, in Table~\ref{tab:result}. 
The critical couplings are here consistent among all the fits, and the 
correlation length exponent is marginally consistent (within 2--3 standard 
deviations).

As seen in the table, The highest relative precision of $g_{\text{c}}$ is obtained 
using $Q_2$ for the full bilayer and $\rho_{\text{s}}$ for the incomplete bilayer.
The latter can probably be traced to very small subleading corrections, as is evident 
already in Fig.~\ref{fig:rho-crss}. For the full bilayer $\rho_{\text{s}}$ also has 
smaller subleading corrections than $Q_2$, but still we obtain a higher accuracy 
in $g_{\text{c}}$ using $Q_2$. This shows that small subleading corrections are
not necessarily advantageous; what matters is how well those corrections are
described by the finite-size scaling forms used. This in turn should be determined 
by the extent of corrections of even higher order. 

For the subleading exponents $\omega$ and $\phi$ we obtain values around unity, 
except in the case of the $\rho_{\text{s}}$ scaling of the incomplete bilayer, 
which requires larger values. For the full bilayer, $\omega = 1.14(3)$, $\phi = 0.8(2)$ 
in the $Q_2$ scaling and $\omega = 1.0(3)$, $\phi = 1.2(2)$ in the $\rho_{\text{s}}$ 
scaling. For the incomplete bilayer, $\omega = 1.0(4)$, $\phi = 1.3(2)$ in $Q_2$ and
$\omega = 1.9(2)$, $\phi = 1.8(2)$ in $\rho_{\text{s}}$. All these subleading
exponents should be interpreted as effective ones, as they are to some extent 
affected by the higher-order corrections that we have neglected. The fact that 
$\omega$ and $\phi$ obtained from $\rho_{\text{s}}$ of the incomplete bilayer are 
close to $2$ suggests that in this case the leading corrections are small and the 
exponents instead reflect predominantly the corrections of the next higher order.

\begin{table}
 \caption{ \label{tab:result}
Bootstrap averages and errorbars for the parameters $g_{\text{c}}, \nu, q_0$ (the values
of the respective quantities at $g_{\text{c}}$). The top group of values are for the incomplete 
bilayer, and the bottom group for complete bilayer. The indicated error bars correspond 
to one standard deviation of the probability distributions obtained in the 
bootstrap analysis, as explained in the text.}
\begin{ruledtabular}
\begin{tabular}{@{\quad}llll}
 & $g_{\text{c}}$ & $\nu$ & $q_0$ \tabularnewline \hline
$Q_2$ & 1.38885(5)  & 0.708(2)& 1.293(3)\tabularnewline
$\rho_{\text{s}} L$& 1.38882(2)  & 0.705(2) & 0.434(3)  \tabularnewline
$\chi_uL$ &  1.388(1) & 0.7(1) & 0.28(2) \tabularnewline \hline
$Q_2$ & 2.52180(3)  & 0.715(2) & 1.2858(3)\tabularnewline
$\rho_{\text{s}} L$& 2.5221(2)  & 0.714(5) & 1.13(3) \tabularnewline
$\chi_uL$ &  2.521(1) & 0.7(1) &  0.12(2)
\end{tabular}
\end{ruledtabular}
\end{table}

\section{\label{summary}Summary and Conclusions}

We have carried out finite-size scaling analyses of high-precision stochastic 
series expansion QMC data for two $S=1/2$ Heisenberg  bilayer models. Using three
different quantities, the Binder order parameter moment ratio, the spin stiffness,
and the long-wavelength magnetic susceptibility, we obtained very accurate
estimates for the critical couplings. We have stressed the importance of
including subleading corrections in the finite-size scaling analysis. All the 
quantities considered then give mutually consistent results for the critical 
couplings as well as for the correlation-length exponent $\nu$. We have assumed 
that the dynamic exponent $z=1$, and all our results are completely consistent 
with this expectation.\cite{ChakravartyPRL1988}

The inclusion of two different subleading corrections in the data fits implies 
larger statistical fluctuations compared to an analysis neglecting subleading 
corrections or taking them into account less completely than we have done here. 
However, because of these procedures, along with careful studies of the 
dependence on the range of system sizes and the data window around the critical 
point, we are confident that the remaining errors are purely statistical and 
accurately captured by the error bars quoted in Table~\ref{tab:result}.
We also note that the different 
quantities we have considered correspond to averaging functions of very different 
properties of the configurations generated in the simulations---the staggered 
magnetization in the case of the Binder ratio, the winding number in the case 
of the spin stiffness, and the long-wavelength magnetization in the case of 
the susceptibility. The consistency among all the results obtained also contribute 
to our confidence in the procedures.

Our final estimates for the critical couplings, taking statistically weighted
averages of the values listed in Table~\ref{tab:result}, are $g_{\text{c}}=2.52181(3)$
and $1.38882(2)$ for the full and incomplete bilayer, respectively. Thus we 
have improved the precision by two orders of magnitude relative
to previous calculations. Knowledge of the critical couplings to this level of
accuracy  should be useful for studies of various aspects of quantum criticality 
at low temperature in these systems, as one can avoid, to a higher degree than
previously, effects from the eventual cross-over to renormalixed classical or 
quantum-disordered behavior as deviations from $g_{\text{c}}$ become relevant as $T\to 0$. 

For the correlation
length, a weighted average of the four results from $\rho_{\text{s}}$ and $Q_2$ listed 
in Table~\ref{tab:result} gives $\nu = 0.7106(9)$. This is consistent within 
error bars with the currently most accurate estimate of the 3D classical Heisenberg 
exponent, $\nu = 0.7112(5)$, obtained from classical 3D Heisenberg simulations
in Ref.~\onlinecite{CompostriniPRB2002}. 

Taking a statistical average of the four individual results for $\nu$ 
(and analogously for $g_{\text{c}}$ discussed above) is motivated here in spite 
of the fact that they are obtained in only two different simulations. 
This is because the winding numbers (giving $\rho_{\text{s}}$) and the sublattice 
magnetization (giving $Q_2$) are very weakly correlated in the simulations. 
We regard it as pure coincidence that the two values for each model in
Table~\ref{tab:result} are closer to each other than those of different models, 
and not an indication of a potentially different universality class in the 
incomplete bilayer (due to incomplete cancelation of Berry phases 
\cite{HaldanePRL1988,ChakravartyPRL1988,ChubukovPRB1994,TroyerJPJ1997})
when the layer-exchange symmetry is not present. The very close agreement
of the universal Binder ratio at $g_{\text{c}}$ also speaks in favor of the same
universality class for both lattices.

Although we have not quite reached the accuracy for $\nu$ obtained in the
most recent classical simulations \cite{CompostriniPRB2002} (although our final 
error bar is actually only approximately twice as large), the precision is still 
sufficiently high to further increase the confidence in the belief that the 
universality class of the transition is that of the 3D classical 
Heisenberg model. The previously best (to our knowledge) determination of $\nu$ 
for the transition in a 2D Heisenberg system  is $0.70(1)$.\cite{MatsumotoPRB2001}

We do not find the predicted \cite{WallinPRB1994}
universality of $\rho_{\text{s}}L$ at the critical
point (whereas we do find the expected universality in the case of the Binder 
ratio); the values for the full and incomplete bilayer differ by about $30\%$. 
On the other hand, in recent simulations of the finite-$T$ transition 
of the 3D $S=1/2$ Heisenberg ferromagnet and antiferromagnet,\cite{heis3d}
we find consistent values for both models at $T_\text{c}$. Thus we are lead to 
speculate that there is some geometrical effect affecting the stiffness, 
so that universality might hold for different critical points (arrangement 
of couplings allowing a transition) on the same lattice but not necessarily
for lattices with different unit cells.

This work was supported by the NSF under grant No.~DMR-0513930.

\end{document}